\documentclass[showpacs,preprintnumbers,amsmath,amssymb]{revtex4}
\usepackage{amsmath,amssymb,graphics,epsfig,subfigure}
\usepackage{color}

\begin{document}
\renewcommand{\baselinestretch}{1.3}

\title{New insights into thermodynamics and microstructure of AdS black holes}

\author{Shao-Wen Wei \footnote{weishw@lzu.edu.cn},
Yu-Xiao Liu \footnote{liuyx@lzu.edu.cn, corresponding author}}

\affiliation{Institute of Theoretical Physics $\&$ Research Center of Gravitation, Lanzhou University, Lanzhou 730000, China}

\maketitle

Black hole, a strange object of extreme gravity, is gradually believed to be a thermodynamic system after 1972. The area and the surface gravity of black hole event horizon were found to be closely related with the entropy and Hawking temperature. Then four thermodynamic laws were established \cite{Bardeen}. Black hole phase transition as a characteristic nature and microstructure indicator has attracted a lot of attention.

An isolated Schwarzschild black hole has negative heat capacity, thus it cannot exist stably. When the charge or angular momentum is considered, a branch of stable large black holes with positive heat capacity will emerge and they can exist stably. Hawking and Page found that there is a phase transition between large black hole and thermal radiation in an AdS space with a negative cosmological constant \cite{Hawking}. Motivated by the AdS/CFT \cite{Maldacena}, this Hawking-Page phase transition was interpreted as the confinement/deconfinement phase transition in gauge theory.

Although there are different kinds of phase transitions, one remarkable difference between a black hole system and an ordinary thermodynamic system is that no pressure term exists for the black hole. A significant step toward this issue was shown in \cite{Kastor}, where the cosmological constant is interpreted as the thermodynamic pressure. Then an AdS black hole system is more like an ordinary thermodynamic system. The corresponding conjugate quantity is the thermodynamic volume, which was found to equal to the geometric volume. However, they have different meanings, and this property are not required to hold for every black hole. Such phase space with the pressure and volume being included is now generally referred to \emph{extended phase space}.

Recently, the $P$-$V$ criticality for charged AdS black holes was studied \cite{Kubiznak}. The results display that a first-order phase transition can occur among the stable small and large black holes at low temperature. At a certain temperature, this phase transition becomes a second-order one, which corresponds to the critical point. At this special point, the critical exponents and scaling laws were calculated. Moreover, they share the same values and forms as that of the van der Waals (VdW) fluid. Therefore, the precise analogy of the phase transition between the AdS black hole and the VdW fluid was established. Subsequently, this study was extended to other black holes, and it is known that the subject is \emph{Black Hole Chemistry}.

Besides the VdW-like phase transition, more interesting phase transitions were found. For example, six-dimensional single spinning Kerr-AdS black hole admits a reentrant phase transition \cite{Altamirano} and multi-spinning Kerr-AdS black hole has a triple point \cite{Mann}. We observed the similar triple point in Gauss-Bonnet gravity, and developed a novel way to shrink the parameter range where the triple point exists by counting the number of the extremal point of the isotherm.

Comparing with the VdW fluid, black hole thermodynamics has two main differences: i) The relation between the black hole thermodynamic volume and specific volume is not linear. This particular property leads to that the equal area law is invalid in the pressure-specific volume diagram for the black hole system. The result motivates us to consider the equal area law in other parameter diagrams. By using them, the phase transition point can be obtained. ii) In some black hole backgrounds, the coexistence curve of the small and large black holes has an analytical form. This allows us to examine the phase transition analytically. The exact values of the critical exponents strongly support the mean field theory. Nevertheless, there is no analytical expression of the coexistence curve for most black hole backgrounds. We can obtain the fitting formula of that curve by numerical calculation just like that of the VdW fluid. This provides a great convenience on exploring properties of black hole systems across the phase transition.

Among the study of the black hole phase transition, more efforts focus on the charged AdS black holes. In fact, the rotating AdS black holes also exhibit phase transitions. However, due to the reason that the state equation is a coupled high power equation of the pressure and volume, the value of the critical point was only obtained in the small angular momentum limit. In order to solve this problem, we reconsidered the means of the critical point \cite{Wei6}. The thermodynamic parameters are divided to two class, the universal parameters and the characteristic parameters. The universal parameters, such as the temperature, pressure, and volume, describe the universal properties of the system. While the characteristic parameters describe the characteristic properties, which includes the charge, angular momentum, and so no. After such a classification, the critical point can be naturally interpreted as the relation between the universal and characteristic parameters. Further combining with the dimensional analysis, the analytical and exact critical points were obtained. Moreover, the starting and ending points of reentrant phase transition were similarly determined \cite{Wei6}.

Further study on this subject was made by considering the microstructure of the black holes. As we know, the black hole entropy is proportional to the event horizon area rather than its volume. The concept of holography theory of gravity originates from this point. Different theories and models aim to understand this particular property. By counting the number of microstates of a weakly coupled D-brane system and then extrapolating the result to the black hole phase, string theory provides a natural framework, and Strominger and Vafa obtained the exact Bekenstein-Hawking entropy formula for certain supersymmetric black holes \cite{Strominger}. Other approaches on calculating the black hole entropy are generally based on the conjecture that gravity is dual to a gauge theory or a strongly coupled conformal field theory. Another interesting model is the fuzzball model \cite{Lunin,Mathur0502050}, which suggests that a black hole is constituted by strings inside the event horizon. Bound states in string theory expand to a size, and the area of the `fuzzball' boundary satisfies a Bekenstein type relation $A/4G\sim S$. The black hole interior is nontrivial and the horizon is not a region of `empty space'. Information in non-BPS states can be carried out in the radiation from the fuzzball boundary. Thus, information paradox and the singularity problem can be well solved in this model.

Despite the great success, each theory or model has its own shortcomings. The question whether a black hole has microscopic structure is still very puzzling. Nevertheless, one can hold the great idea of Boltzmann that ``If you can heat it, it has microscopic structure". This definitely confirms that a black hole must have microstructure because its temperature can be changed by emitting or absorbing particles. Phase transition provides a potential convenience to insight the black hole microstructure because the microstructure of a thermodynamic system greatly changes when it encounters a first-order phase transition. The quantity, number density, is a key to investigate the change of the physical properties of the black hole microstructure during the phase transition. Therefore, we assumed that a black hole is constituted by some unknown microscopic molecules \cite{Weiw}. After making a comparison with the vdW fluid, we introduced a new concept of number density for the black hole microstructure, with which change of the black hole property across the phase transition was well understood.

Very recently, a big success was obtained by applying the geometric approach. Ruppeiner geometry \cite{Ruppeiner} constructing from the fluctuation theory provides us a significant tool on understanding the microstructure of a thermodynamic system. The line element and scalar curvature of the geometry have the following interpretation: i) The less probable a fluctuation between two thermodynamic states, the further apart the distance measuring by the geometry are. ii) Negative (positive) scalar curvature indicates that an attractive (a repulsive) interaction dominates, whereas vanishing curvature signifies repulsive and attractive interactions are in balance. iii) Moreover, we conjecture that the absolute value of the scalar curvature is related to the strength of the interactions among the microstructures.

\begin{figure}
\includegraphics[width=6cm]{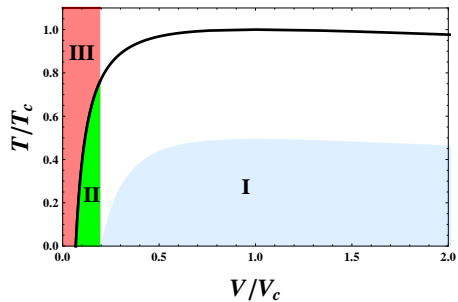}
\caption{Behavior of the normalized scalar curvature in the $T/T_{c}-V/V_{c}$ (temperature-volume) diagram with $T_{c}$ and $V_{c}$ being the critical temperature and volume of the phase transition. In regions I, II, and III the corresponding curvature is positive indicating a repulsive interaction. Otherwise, it is negative, which implies a attractive interaction dominates the system. The black line denotes the coexistence curve.}\label{ppss}
\end{figure}

For the VdW fluid, the result shows that the attractive interaction is dominant among the molecules. However, when we apply the approach to the charged AdS black hole, the geometry becomes problematic due to the vanishing heat capacity at constant volume. A natural idea is that we can treat the black hole system as a special thermodynamic in the limit of $k_B\rightarrow 0^+$. Taking this new interpretation, we proposed a new normalized scalar curvature \cite{Weiwe}. Here we show the corresponding scalar curvature in Fig. \ref{ppss}. In regions I, II, and III, the positive curvature implies that the repulsive interaction dominates. However, considering that the curvature is calculated by using the equation of state, while the equation of sate may not hold in the coexistence region. So regions I and II should be excluded due to this reason. Nevertheless, the repulsive interaction is still allowed in region I. We observe a repulsive interaction for the small black hole at high temperature despite the black hole charge. This result strongly suggests that, the charged AdS black hole has a significant different interaction between the microstructures from the VdW fluid although they have the similar phase transition. The normalized scalar curvature was also found to go to negative infinity near the critical point. Moreover a universal critical exponent 2 was obtained. For the higher-dimensional charged AdS black holes, the results indicate that the black hole microstructures are quite similar but have a tiny difference.

An intriguing microstructures were observed for the five-dimensional neutral Gauss-Bonnet (GB) AdS black hole \cite{Wei1910.04528}. In this background, there exists a small-large black hole phase transition. When the system crosses the coexistence curve, the microstructures will occur an obvious change. However, the scalar curvature of the corresponding Ruppeiner geometry is the same for the saturated small and large black holes. This novel property indicates that the black hole microstructures keep unchange when the system crosses the coexistence curve. It provides us a first interesting example towards the modified gravity.

In summary, black hole thermodynamics and phase transition can provide us a significant approach to investigate the black hole microstructures. Combining with the geometrical technique, the property and interaction of the microstructures can be obtained. This treatment is completely new and worth to generalized to other black hole backgrounds. These will insight into the nature of black hole microstructures and gravity.

\section*{Acknowledgements}
 This work was supported in part by the National Natural Science Foundation of China (Grants No. 11675064 and No. 11875151).

\end{document}